\documentclass[a4paper,UKenglish,cleveref, autoref, thm-restate]{lipics-v2021}

\pdfoutput=1 
\hideLIPIcs  

\title{Is Self-Admitted Technical Debt Tested? An Empirical Study of Coverage, Co-change, and Impact}
\titlerunning{Is Self-Admitted Technical Debt Tested?} 

\author{Suzuka {Yoshimoto}}{Nara Institute of Science and Technology, Japan}{}
{https://orcid.org/0009-0006-5366-8014}{}

\author{Kosei {Horikawa}}{Nara Institute of Science and Technology, Japan}{}
{https://orcid.org/0009-0006-6317-0754}{}

\author{Daniel {Feitosa}}{University of Groningen, the Netherlands}{}
{https://orcid.org/0000-0001-9371-232X}{}

\author{Yutaro {Kashiwa}}{Nara Institute of Science and Technology, Japan}{yutaro.kashiwa@naist.ac.jp}{https://orcid.org/0000-0002-9633-7577}{}

\author{Hajimu {Iida}}{Nara Institute of Science and Technology, Japan}{}
{https://orcid.org/0000-0002-2919-6620}{}

\authorrunning{S. Yoshimoto and K. Horikawa and D. Feitosa and Y. Kashiwa and H. Iida}

\Copyright{Suzuka Yoshimoto and Kosei Horikawa and Daniel Feitosa and Yutaro Kashiwa and Hajimu Iida} 



\category{Technical Track Paper}

\relatedversion{} 

\nolinenumbers 

\EventEditors{Robert Feldt, Maria Paasivaara, Daniel Mendez, Stefan Wagner, and Marvin Mu\~{n}oz Bar\'{o}n}
\EventNoEds{5}
\EventLongTitle{20th International Symposium on Empirical Software Engineering and Measurement (ESEM 2026)}
\EventShortTitle{ESEM 2026}
\EventAcronym{ESEM}
\EventYear{2026}
\EventDate{October 8--9, 2026}
\EventLocation{Munich, Germany}
\EventLogo{}
\SeriesVolume{394}
\ArticleNo{36}

\usepackage{fancyvrb}
\usepackage{xurl}
\usepackage{booktabs}
\usepackage{url}
\usepackage{amsmath,amsfonts}
\usepackage{algorithmic}
\usepackage{graphicx}
\usepackage{textcomp}
\usepackage{xspace}
\usepackage{subcaption}
\usepackage{tabularx}
\usepackage{framed}
\usepackage{chngpage}
\usepackage[many]{tcolorbox}
\usepackage{pgfplots}
\pgfplotsset{compat=newest}
\usepackage{siunitx}
\usepackage{balance}
\usepackage{enumitem}
\usepackage{listings} 
\usepackage{multirow} 
\usepackage{xfp}      

\usepackage{hyperref}
\usepackage{threeparttable}

\usepackage{tikz}

\usepackage{xcolor}
\definecolor{commentgreen}{rgb}{0, 0.4, 0} 
\definecolor{keywordblue}{rgb}{0, 0, 0.8}  

\lstdefinestyle{javastyle}{
    backgroundcolor=\color{backcolour},   
    commentstyle=\color{codegreen},
    keywordstyle=\color{magenta},
    numberstyle=\tiny\color{codegray},
    stringstyle=\color{codepurple},
    basicstyle=\ttfamily\small,
    breakatwhitespace=false,         
    breaklines=true,                 
    captionpos=t, 
    abovecaptionskip=10pt, 
    belowcaptionskip=10pt, 
    frame=single, 
    numbers=left,                    
    numbersep=5pt,                  
    showspaces=false,                
    showstringspaces=false,
    showtabs=false,                  
    tabsize=2,
    language=Java
}

\newcommand{\rqa}{$RQ_1$}
\newcommand{\rqb}{$RQ_2$}
\newcommand{\rqc}{$RQ_3$}
\newcommand{\rqaa}{To what extent is SATD covered by tests compared to the overall codebase?}
\newcommand{\rqbb}{To what extent do developers add targeted tests when removing SATD?}
\newcommand{\rqcc}{What is the impact of removing SATD without adequate testing on software quality?}

\newcommand{\rqA}{\rqa: \rqaa}
\newcommand{\rqB}{\rqb: \rqbb}
\newcommand{\rqC}{\rqc: \rqcc}

\newcommand{\lang}{{\sc Commons Lang}\xspace}
\newcommand{\io}{{\sc Commons IO}\xspace}
\newcommand{\hibernate}{{\sc Hibernate}\xspace}
\newcommand{\dubbo}{{\sc Dubbo}\xspace}
\newcommand{\jfree}{{\sc JFreeChart}\xspace}
\newcommand{\spoon}{{\sc Spoon}\xspace}
\newcommand{\maven}{{\sc Maven}\xspace}
\newcommand{\storm}{{\sc Storm}\xspace}

\newcommand{\TotalFuzzingProjectNum}{1311}
\newcommand{\TargetProjectNum}{628}

\pgfmathsetmacro{\TargetProjectRateNum}{round((\TargetProjectNum/\TotalFuzzingProjectNum)*100*10)/10}

\newcommand{\ExcludedProjectNum}{4}
\pgfmathsetmacro{\FinallyTargetProjectInt}{round(\TargetProjectNum - \ExcludedProjectNum)}

\usepackage{fp}
\DeclareRobustCommand{\calcpct}[2]{%
  \FPeval{\rawres}{(#1 / #2) * 100}
  \FPround{\finalres}{\rawres}{1}
  \finalres
} 

\definecolor{darkgreen}{rgb}{0, 0.5, 0} 
\definecolor{whitesmoke}{rgb}{0.99, 0.99, 0.99} 

\def\Underline{\setbox0\hbox\bgroup\let\\\endUnderline}
\def\endUnderline{\vphantom{y}\egroup\smash{\underline{\box0}}\\}

\newcommand{\eg}{\textit{e.g.,}\xspace}

\newcounter{findings_no}

\usepackage{listings}
\definecolor{backcolour}{rgb}{0.95,0.95,0.92}
\lstdefinelanguage{diff}{
  morecomment=**[f][\color{red}]{-},         
  morecomment=**[f][\color{darkgreen}]{+},       
  moredelim=**[is][\bfseries]{@@}{@@},
}
\definecolor{backcolour}{rgb}{0.95,0.95,0.92}
\lstdefinelanguage{commit}{ 
  breakindent = 0pt,
  numbers=none,
  backgroundcolor=\color{white},
  frame=single,
  xleftmargin=3.5em,
  numbersep=0em,
  xrightmargin=1.5em,
}

\usepackage[many]{tcolorbox}  
\tcbuselibrary{listings,breakable}
\definecolor{main}{HTML}{D0D3D4}    
\definecolor{sub}{HTML}{D0D3D4}     
\newtcolorbox{dbox}{
    left=2pt,right=2pt,top=2pt,bottom=2pt,
    enhanced, 
    boxrule = 0pt,
    enlarge top by=5pt,
    enlarge bottom by=3pt,
  }
\tcbset{
    sharp corners,
    before skip = 0.1cm,    
    after skip = 0.01cm      
}

\def\summarybox#1#2{
\medskip
\begin{tcolorbox}[
  enhanced,
  title=#1,
  colframe=darkgray,
]
    #2
\end{tcolorbox}
}

\AtBeginDocument{%
  \providecommand\BibTeX{{%
    \normalfont B\kern-0.5em{\scshape i\kern-0.25em b}\kern-0.8em\TeX}}}

\begin{document}






\maketitle
\begin{abstract}
\textbf{Background.} When developers write a \texttt{TODO} or \texttt{FIXME} comment, they are explicitly admitting that the code is suboptimal: a built-in warning that this logic deserves extra scrutiny. Yet it is an open question whether Self-Admitted Technical Debt (SATD) actually receives that scrutiny in the form of software testing.
\textbf{Aim.} We aim to characterize the relationship between SATD and testing across three dimensions: the extent to which SATD-affected code is covered by existing tests, whether developers synchronize test additions with debt resolution, and whether such testing affects the long-term observability of resulting defects.
\textbf{Method.} For that, we conducted an empirical study on eight open-source Java projects, analyzing test coverage of 784 SATD instances identified in the latest releases and performing a longitudinal examination of \num{5175} SATD removal events.
\textbf{Results.} Our results show that while 60.7\% of SATD-affected code is covered by existing test suites, developers rarely synchronize test modifications with debt resolution; manual inspection confirms that only 3.4\% of SATD removal commits include new tests specifically targeting the resolved debt (vs.\ 12.5\% that co-add tests in the same commit). Longitudinal analysis further suggests that SATD resolutions exhibit nearly identical localized bug induction rates within short-to-medium-term windows regardless of test modifications. However, over a longer, unrestricted observation window, a slight divergence emerges where the test-added group reaches a higher cumulative defect alignment probability (\num{6.32}\% vs.\ \num{4.37}\%), a counterintuitive trend potentially driven by the selective testing of inherently complex components.
\textbf{Conclusion.} Developers treat SATD repayment as an ordinary code change rather than as a high-risk maintenance activity: most debt removals proceed without targeted verification, despite the developer's own prior flag that the code is suboptimal. 
\end{abstract}



\section{Introduction}\label{sec:introduction}
Technical debt is an inherent byproduct of software development: facing deadlines or incomplete specifications, developers sometimes implement solutions they know are suboptimal \cite{DBLP:journals/software/LimTS12}. Rather than leaving that knowledge implicit, they often record it directly in the source code through comments such as \texttt{TODO}, \texttt{FIXME}, and \texttt{HACK}. This practice, known as Self-Admitted Technical Debt (SATD)~\cite{DBLP:conf/icsm/PotdarS14}, is surprisingly widespread: between 2.4\% and 31\% of source files in open-source projects contain such comments~\cite{DBLP:conf/icsm/PotdarS14}, and the flagged code tends to persist for tens to hundreds of days before being resolved~\cite{DBLP:conf/saner/RecupitoMNP25, DBLP:journals/corr/abs-2311-12019}, during which time it continues to be more change-prone and defect-prone than surrounding code~\cite{DBLP:conf/wcre/WehaibiSG16}.

What makes SATD distinctive is that the developer's own comment serves as an explicit warning: this code is known to be fragile, incomplete, or incorrect. Unlike latent defects that remain hidden until discovered, SATD represents a conscious acknowledgment that the implementation may not behave as intended. Such code should therefore be subject to verification, and software testing is the primary mechanism through which developers confirm that code behaves correctly and does not regress under future changes.

Software testing serves as a primary mechanism for early defect detection and regression prevention. A rich body of literature has examined the relationship between test coverage and defect detection capability~\cite{DBLP:conf/sigsoft/IvankovicPJF19}, revealing, for example, that higher statement or branch coverage correlates with greater fault-detection effectiveness, though the strength of this relationship varies across projects and coverage metrics~\cite{DBLP:conf/qsic/KochharBLJ13}. Researchers have also studied how test modifications affect code quality, finding that commits accompanied by test changes tend to introduce fewer regressions than those without~\cite{DBLP:conf/wcre/Wang000W21}. More broadly, tests act as a safety net during refactoring and code modifications, and the presence or absence of adequate tests can influence the success of such changes~\cite{DBLP:journals/ese/ZaidmanRDD11}.

Despite these parallel lines of research, SATD studies and testing studies have progressed largely in isolation. These findings are naturally complementary, yet the relationship between SATD and testing remains largely unexplored. First, it is unknown to what extent code containing SATD is protected by test suites. Second, there is a lack of empirical evidence on how often commits that remove SATD are accompanied by changes to test code, that is, the extent to which developers are conscious of testing during the SATD lifecycle. Third, whether adding tests at the time of SATD removal actually reduces the rate of subsequent bug occurrences has not been verified.

To address these gaps, we conducted an empirical study on eight open-source Java projects hosted on GitHub, examining the relationship between SATD and testing across two temporal scales. 
First, we identified 784 SATD instances in the latest releases to evaluate their current test coverage. 
Second, we performed a longitudinal analysis of \num{5175} SATD removal events to investigate the co-occurrence of test modifications and the long-term observability of resulting defects. 
Based on this extensive dataset, we pose the following three research questions and summarize our key findings:

\smallskip\noindent\textbf{\rqA}
We found that 60.7\% of SATD-affected code is covered by test cases. 
Furthermore, when categorizing SATD into technical types (e.g., Design, Requirement, and Defect debt), median coverage remained similar across types (41.7\%--50.0\%), indicating that the specific type of debt does not significantly influence whether it gets tested. 
These results indicate that a majority of admitted debt is within the reach of existing verification frameworks. Notably, in several projects, SATD-affected code exhibited higher test coverage compared to the overall codebase, indicating that developers do not necessarily neglect brittle areas. 

\smallskip\noindent\textbf{\rqB}
Despite the high baseline coverage, our analysis of \num{5175} removed SATD revealed a disconnect during the debt lifecycle; only 3.4\% of SATD removals included concurrent modifications to test code. This suggests that while developers may rely on a pre-existing safety net, they rarely update or tailor tests specifically in response to the resolution of known debt.

\smallskip\noindent\textbf{\rqC}
To understand the impact of these rare testing efforts, we analyzed the relationship between bug induction and SATD removals that included concurrent test additions. Our results indicate that the presence of test modifications yields no immediate reduction in localized bug induction rates within short-to-medium-term windows (\eg 30, 90, and 180 days). Interestingly, over an unrestricted time window, a slight divergence becomes visible, with the test-added group exhibiting a higher cumulative probability of subsequent defect alignment than the no-test group. This counterintuitive pattern suggests that general test modifications within a commit do not inherently prevent regressions, potentially because they lack sufficient coverage for the resolved logic. Alternatively, this may reflect a selection bias where developers selectively implement tests for particularly complex or critical SATD instances that remain inherently prone to defects. 


\section{Motivating Example}\label{sec:motivation}

Software maintenance requires a continuous, synchronized effort between the evolution of production code and its corresponding test suite. 
Prior studies on code co-evolution, such as the work by \textit{Levin and Yehudai} \cite{DBLP:conf/icsm/LevinY17}, have observed that this synchronization is often neglected in practice; developers frequently perform code fixes without correspondingly updating or adding tests within the same commit. 
While such findings highlight a general disconnect in maintenance activities, a critical question remains: \textit{Does this trend persist even when developers have explicitly flagged the code as fragile or suboptimal?}

When developers write an SATD comment like \texttt{TODO} or \texttt{FIXME}, they are explicitly admitting that the logic is suboptimal---a built-in warning that the code deserves extra scrutiny. 
In an ideal development workflow, the resolution of such debt should be a controlled, verified process. 
For instance, in the Dubbo project (Listing~\ref{lst:satd_before}), a developer identified a limitation where the system only checked a single registry. 
When this debt was eventually repaid (Listing~\ref{lst:satd_after}), the developer not only refactored the logic but also provided a corresponding test case (Listing~\ref{lst:satd_test}) to verify the new implementation. 
This scenario illustrates a rigorous approach where the removal of admitted debt is immediately fortified by quality assurance.

\begin{lstlisting}[
    float=hbt!,
    language=Java,
    basicstyle=\ttfamily\small,
    commentstyle=\color{commentgreen},
    keywordstyle=\color{keywordblue},
    breaklines=true,
    breakatwhitespace=false,
    % postbreak=\mbox{\textcolor{gray}{$\hookrightarrow$}\space},
    captionpos=t,
    abovecaptionskip=10pt,
    belowcaptionskip=12pt,
    caption={Original code with SATD (\dubbo, \texttt{ServiceCheckUtils.java})},
    label={lst:satd_before}
]
public static boolean isRegistered(ProviderModel providerModel) {
    // TODO, only check the status of one registry and no protocol now.
    Collection<Registry> registries = registryManager.getRegistries(); // setup omitted
    if (CollectionUtils.isNotEmpty(registries)) {
        AbstractRegistry abstractRegistry = (AbstractRegistry) registries.iterator().next();
        // ... omitted for brevity ...
    }
}
\end{lstlisting}

\begin{lstlisting}[
    float=hbt!,
    language=Java,
    basicstyle=\ttfamily\small,
    commentstyle=\color{commentgreen},
    keywordstyle=\color{keywordblue},
    breaklines=true,
    breakatwhitespace=true,
    captionpos=t,
    abovecaptionskip=10pt,
    belowcaptionskip=12pt,
    caption={Refactored code after SATD removal (\dubbo, \texttt{ServiceCheckUtils.java})},
    label={lst:satd_after}
]
public static boolean isRegistered(ProviderModel providerModel) {
    // check all registries status
    for (ProviderModel.RegisterStatedURL registerStatedURL : providerModel.getStatedUrl()) {
        if (registerStatedURL.isRegistered()) {
            return true;
        }
    }
}
\end{lstlisting}

\begin{lstlisting}[
    float=hbt!,
    language=Java,
    basicstyle=\ttfamily\small,
    commentstyle=\color{commentgreen},
    keywordstyle=\color{keywordblue},
    breaklines=true,
    breakatwhitespace=true,
    captionpos=t,
    abovecaptionskip=10pt,
    belowcaptionskip=12pt,
    caption={Newly added test code accompanying the SATD removal (\dubbo, \texttt{ServiceCheckUtilsTest.java})},
    label={lst:satd_test}
]
@Test
public void testIsRegistered() {
    // ... test setup omitted for brevity ...
    boolean registered = ServiceCheckUtils.isRegistered(providerModel);
    assertFalse(registered);
}
\end{lstlisting}

\begin{lstlisting}[
    float=hbt!,
    language=Java,
    basicstyle=\ttfamily\small,
    commentstyle=\color{commentgreen},
    keywordstyle=\color{keywordblue},
    breaklines=true,
    breakatwhitespace=false,
    captionpos=t,
    abovecaptionskip=10pt,
    belowcaptionskip=12pt,
    caption={Original code with SATD (\hibernate, \texttt{DenormalizedTable.java})},
    label={lst:hibernate_before}
]
@Override
public Iterator getUniqueKeyIterator() {
    //wierd implementation because of hacky behavior
    //of Table.sqlCreateString() which modifies the
    //list of unique keys by side-effect on some
    //dialects
    Map uks = new HashMap();
    uks.putAll( getUniqueKeys() );
    uks.putAll( includedTable.getUniqueKeys() );
    return uks.values().iterator();
}
\end{lstlisting}

\begin{lstlisting}[
    float=hbt!,
    language=Java,
    basicstyle=\ttfamily\small,
    commentstyle=\color{commentgreen},
    keywordstyle=\color{keywordblue},
    breaklines=true,
    breakatwhitespace=false,
    captionpos=t,
    abovecaptionskip=10pt,
    belowcaptionskip=12pt,
    caption={Bug-inducing code: SATD removed without accompanying test additions (\hibernate, \texttt{DenormalizedTable.java})},
    label={lst:hibernate_buggy}
]
@Override
public Iterator getUniqueKeyIterator() {
    Iterator iter = includedTable.getUniqueKeyIterator();
    while ( iter.hasNext() ) {
        UniqueKey uk = (UniqueKey) iter.next();
        createUniqueKey( uk.getColumns() );
    }
    return getUniqueKeys().values().iterator();
}
\end{lstlisting}

However, the presence of a \texttt{TODO} or \texttt{FIXME} does not always act as a trigger for such caution. 
Another scenario is illustrated by the Hibernate project. 
In Listing~\ref{lst:hibernate_before}, a developer acknowledged a ``wierd implementation'' caused by ``hacky behavior.'' 
While the SATD comment was later removed in an attempt to clean up the logic (Listing~\ref{lst:hibernate_buggy}), this change was not accompanied by any new tests. 
Without a safety net, this testless debt removal introduced a regression that was later identified as a bug-inducing change.

This discrepancy reveals a blind spot in technical debt management. 
While prior research has focused heavily on the identification and classification of SATD, there is a lack of systematic understanding regarding the testing. 
Even if general code changes often lack tests, the failure to verify code that has been explicitly admitted as problematic represents a higher-risk category of neglect. 
This study aims to bridge this gap by quantifying the relationship between testing practices and the SATD lifecycle, providing the empirical link between these two previously disconnected domains of software engineering research.

\newcommand{\numCommonsLangComments}{1042}
\newcommand{\numCommonsLangSATD}{28}

\newcommand{\numCommonsIOComments}{568}
\newcommand{\numCommonsIOSATD}{23}

\newcommand{\numJFreeChartComments}{2865}
\newcommand{\numJFreeChartSATD}{60}

\newcommand{\numHibernateComments}{11808}
\newcommand{\numHibernateSATD}{530}

\newcommand{\numSpoonComments}{2207}
\newcommand{\numSpoonSATD}{25}

\newcommand{\numDubboComments}{924}
\newcommand{\numDubboSATD}{21}

\newcommand{\numMavenComments}{862}
\newcommand{\numMavenSATD}{69}

\newcommand{\numStormComments}{3563}
\newcommand{\numStormSATD}{43}

\newcommand{\numTotalSATD}{\inteval{%
  \numCommonsLangSATD + \numCommonsIOSATD + \numJFreeChartSATD + \numHibernateSATD + \numSpoonSATD + \numDubboSATD + \numMavenSATD + \numStormSATD
}}
\newcommand{\numTotalComments}{\inteval{%
  \numCommonsLangComments + \numCommonsIOComments + \numJFreeChartComments + \numHibernateComments + \numSpoonComments + \numDubboComments + \numMavenComments + \numStormComments
}}

\newcommand{\numDesignSATD}{518}
\newcommand{\numRequirementSATD}{256}
\newcommand{\numDefectSATD}{22}
\newcommand{\numDocumentationSATD}{0}
\newcommand{\numTestSATD}{3}

\section{Study Design}\label{sec:studydesign}


\subsection{Research Questions}\label{sec:rqs}

The following research questions and their motivations guide our empirical investigation.

\smallskip\noindent\textbf{\rqA}\par
SATD represents a developer's own acknowledgment that a piece of code is suboptimal or incomplete. Prior work has shown that such
debt-ridden code tends to be more change-prone and defect-prone than surrounding code~\cite{DBLP:conf/wcre/WehaibiSG16}, and that SATD comments persist for tens to hundreds of days before being resolved~\cite{DBLP:conf/saner/RecupitoMNP25,
DBLP:journals/corr/abs-2311-12019}. Given these known risks, one might expect developers to compensate with more thorough testing. Yet it is equally plausible that writing tests for SATD-affected code, much like the resolution of the debt itself, is also postponed under time pressure. While prior studies have examined the relationship between test coverage and defect detection~\cite{DBLP:conf/sigsoft/IvankovicPJF19, DBLP:conf/qsic/KochharBLJ13}, no study has directly investigated whether SATD-affected code receives adequate test coverage. We therefore aim to determine the extent to which SATD is covered by test suites and how this coverage varies across
different types of debt.
    
\smallskip\noindent\textbf{\rqB}\par
Static coverage alone cannot tell us whether developers are consciously investing in test quality during the SATD lifecycle, or whether any existing coverage is merely incidental. 
Prior studies on developer behavior during refactoring have found that test updates are often neglected even when production code changes substantially~\cite{DBLP:journals/ese/ZaidmanRDD11}.
Whether this pattern extends to SATD-related commits, where the code is already known to be suboptimal, remains an open question.
Understanding this reveals whether quality assurance is treated as an integral part of debt management or whether it is systematically overlooked during these high-risk changes.
    
\smallskip\noindent\textbf{\rqC}\par
Prior work has shown that SATD-related changes tend to be more complex and harder to perform than typical modifications~\cite{DBLP:conf/wcre/WehaibiSG16}, indicating that debt resolution carries inherent risks.
However, no study has directly quantified how concurrent test additions during SATD removal influence the observability of subsequent defects.
Therefore, we aim to evaluate the effect of concurrent test additions on linked bug-inducing rates during debt resolution, investigating whether such tests serve as an effective diagnostic mechanism for capturing regressions that would otherwise persist as undetected failures.

\subsection{Subject Projects}\label{sec:subject_projects}
For our study, we selected a diverse set of eight open-source Java projects, as detailed in Table~\ref{tab:projects_simple}. While purely static SATD studies analyze larger corpora ~\cite{DBLP:conf/promise/PalmaABML18, 10.1145/3786791}, our design prioritizes dynamic test execution and manual inspection, requiring stable build environments and a deliberate trade-off in corpus size.
The chosen projects vary in domain and scale, covering utility libraries, application frameworks, specialized tools, and build infrastructure.
This allows us to generalize our findings across different software development contexts. 
For each project, we analyzed a specific, recent release to ensure the relevance of our findings. The projects are divided into two categories based on their pre-existing test coverage infrastructure, which is central to our analysis.

Jacoco-preconfigured projects comprises four projects that
already had the Jacoco test coverage plugin configured. This group
includes foundational core utility libraries (\lang and \io) and
large-scale application frameworks (\hibernate, an Object-Relational
Mapping framework, and \dubbo, a microservice framework).

Manually configured projects comprises four projects for
which we manually added the Jacoco plugin to perform our analysis.
This group includes specialized libraries (\jfree, a charting
library, and \spoon, a metaprogramming library) and development
infrastructure tools ({\sc Maven}, a build automation tool, and
\storm, a real-time stream processing system).

\begin{table*}[!t]
\footnotesize 
\centering
\caption{Overview of SATD found within method bodies in analyzed projects.}
\label{tab:projects_simple}
\setlength{\tabcolsep}{4pt}
\begin{tabularx}{\textwidth}{@{} l >{\raggedright\arraybackslash}X >{\raggedright\arraybackslash}X rrr @{}}
\toprule
\textbf{Repository} & \textbf{Release} & \textbf{Module(s)} & \textbf{\#Comments} & \textbf{\#SATD} & \textbf{\% SATD} \\ \midrule
\multicolumn{6}{l}{\textit{Group 1: Projects with preconfigured Jacoco plugin}} \\\addlinespace
~~~~\lang~\cite{repo:commons-lang} & commons-lang-3.18.0 & src & \num{\numCommonsLangComments} & \num{\numCommonsLangSATD} & \calcpct{\numCommonsLangSATD}{\numCommonsLangComments} \\
~~~~\io~\cite{repo:commons-io} & commons-io-2.20.0 & src & \num{\numCommonsIOComments} & \num{\numCommonsIOSATD} & \calcpct{\numCommonsIOSATD}{\numCommonsIOComments} \\
~~~~\hibernate~\cite{repo:hibernate} & 6.6.28 & hibernate-core & \num{\numHibernateComments} & \num{\numHibernateSATD} & \calcpct{\numHibernateSATD}{\numHibernateComments} \\
~~~~\dubbo~\cite{repo:dubbo} & dubbo-3.3.5 & dubbo-common & \num{\numDubboComments} & \num{\numDubboSATD} & \calcpct{\numDubboSATD}{\numDubboComments} \\
\midrule
\multicolumn{6}{l}{\textit{Group 2: Projects where Jacoco plugin was manually added}} \\\addlinespace
~~~~\jfree~\cite{repo:jfreechart} & v1.5.6 & src & \num{\numJFreeChartComments} & \num{\numJFreeChartSATD} & \calcpct{\numJFreeChartSATD}{\numJFreeChartComments} \\
~~~~\spoon~\cite{repo:spoon} & v11.2.1 & src & \num{\numSpoonComments} & \num{\numSpoonSATD} & \calcpct{\numSpoonSATD}{\numSpoonComments} \\
~~~~\maven~\cite{repo:maven} & maven-3.9.11 & maven-core, maven-compat & \num{\numMavenComments} & \num{\numMavenSATD} & \calcpct{\numMavenSATD}{\numMavenComments} \\
~~~~\storm~\cite{repo:storm} & v2.8.2 & storm-server, storm-client & \num{\numStormComments} & \num{\numStormSATD} & \calcpct{\numStormSATD}{\numStormComments} \\
\midrule
\textbf{Total} & & & \textbf{\num{\numTotalComments}} & \textbf{\numTotalSATD} & \calcpct{\numTotalSATD}{\numTotalComments} \\
\bottomrule
\end{tabularx}
\end{table*}

\subsection{Data Collection and Processing Pipeline}\label{sec:data_pipeline}

Our methodology involves a multi-stage pipeline to detect SATD, measure its test coverage, and trace its lifecycle and impact on defects.

\smallskip\noindent\textbf{1) SATD Detection and Classification.}\label{sec:satd_detection}
We employed DebtHunter~\cite{DBLP:conf/ease/SalaTF21}, a state-of-the-art, ML-based SATD detection tool. The study introducing the tool demonstrated its high performance, with a precision of 97.2\%, a recall of 96.7\%, and an F1-score of 96.5\%, outperforming other detection tools. 
In addition to SATD detection, we use DebtHunter to classify SATD by type. Since our study focuses on technical debt within executable logic, we modified the tool for two purposes: (1) to restrict detection to comments within method bodies, and (2) to extract the location of each SATD instance, including file name, method name, and line number. 

To independently verify the reliability of DebtHunter, we manually validated its detection results on the three projects not included in its original training: \dubbo, \spoon, and \storm. We inspected all 129 comments flagged as SATD by the tool across these three projects (prior to applying the method body restriction). Our review confirmed that the classifications for \spoon and \storm were highly accurate, identifying zero false positives. For \dubbo, however, we found 10 false positives, which were non-SATD comments related to configuration (\eg \texttt{load dubbo.applications.xxx}). We excluded these 10 false positives from our dataset before proceeding with the analysis.

Using DebtHunter, we categorized the instances into five types: Design, Requirement, Defect, Test, and Documentation. As shown in Table \ref{tab:satd_classification}, our dataset is dominated by Design (logical structure) and Requirement (requirement specifications) debt.

\begin{table}[!bt]
\small
\centering
\caption{Classification of all \num{\numTotalSATD} SATD instances.}
\label{tab:satd_classification}
\begin{tabular}{lrr}
\toprule
\textbf{Type} & \textbf{\#} & \textbf{\%} \\ \midrule
Design & \num{\numDesignSATD} & \calcpct{\numDesignSATD}{\numTotalSATD} \\
Requirement & \num{\numRequirementSATD} & \calcpct{\numRequirementSATD}{\numTotalSATD} \\
Defect & \num{\numDefectSATD} & \calcpct{\numDefectSATD}{\numTotalSATD} \\
Test & \num{\numTestSATD} & \calcpct{\numTestSATD}{\numTotalSATD} \\
Documentation & \num{\numDocumentationSATD} & \calcpct{\numDocumentationSATD}{\numTotalSATD} \\
\bottomrule
\end{tabular}
\end{table}

\smallskip\noindent\textbf{2) Measuring Test Coverage via Tracer Lines.}\label{sec:coverage_method}
To measure test coverage, we transform the passive SATD comment into an active, executable tracer line. For each identified SATD, we attempt to replace the comment line with an executable statement (i.e., \texttt{System.out.println("SATD: IMPLEMENTATION");}). We then run the project's entire test suite and analyze the generated jacoco.xml reports. An SATD is considered covered if an inserted tracer line is reported as executed by Jacoco.
However, this direct replacement fails in two scenarios, for which we manually analyze the SATD and map it to an appropriate executable line within the relevant debt-ridden code block.

\noindent\textit{(i) Inappropriate Location:} The comment is not located within the actual debt-ridden code block it refers to. For example, in Listing \ref{fig:example-location}, the \texttt{FIXME} comment is placed above the \texttt{if} block, but it refers to the logic within the \texttt{for} loop. Replacing the comment line directly would test the area outside the relevant logic.
\begin{lstlisting}[
    float=hbt!,
    language=Java,
    basicstyle=\ttfamily\small,
    commentstyle=\color{commentgreen},
    keywordstyle=\color{keywordblue},
    breaklines=true,
    breakatwhitespace=false,
    % postbreak=\mbox{\textcolor{gray}{$\hookrightarrow$}\space}, % 改行箇所に記号を入れる（任意）
    captionpos=t,
    abovecaptionskip=0pt,
    belowcaptionskip=12pt,
    caption={Example of an SATD comment in an inappropriate location
    (\hibernate, \texttt{MapBinder.java)}.},
    label={fig:example-location}
]
//FIXME pass the Index Entity JoinColumns
if ( !collection.isOneToMany() ) {
    //index column should not be null
    for ( AnnotatedJoinColumn column : mapKeyManyToManyColumns.getJoinColumns() ) {
        column.forceNotNull();
    }
}
\end{lstlisting}

\bigskip
\noindent\textit{(ii) Syntax Error: The replacement violates Java syntax.} In Listing \ref{fig:example-syntax}, the \texttt{TODO} comment is located within a variable declaration statement. Inserting an executable statement at the comment's location would break the assignment logic, causing a compilation error.
\begin{lstlisting}[
    float=hbt!,
    language=Java,
    basicstyle=\ttfamily\small,
    commentstyle=\color{commentgreen},
    keywordstyle=\color{keywordblue},
    breaklines=true,
    breakatwhitespace=false,
    % postbreak=\mbox{\textcolor{gray}{$\hookrightarrow$}\space}, % 改行箇所に記号を入れる（任意）
    captionpos=t,
    abovecaptionskip=0pt,
    belowcaptionskip=12pt,
    caption={Example of an SATD comment causing a syntax error if replaced 
(\hibernate, \texttt{SqmSelectionQueryImpl.java)}.},
    label={fig:example-syntax}
]
protected List<R> doList() {
    final boolean containsCollectionFetches =
            //TODO: why is this different from QuerySqmImpl.doList()?
            statement.containsCollectionFetches();
    // ... omitted for brevity ...
}
\end{lstlisting}

\smallskip\noindent\textbf{3) Mining SATD Lifecycle and Test Additions.}\label{sec:lifecycle_method}
To detect SATD-removing events, we utilized a customized version of SATDBailiff~\cite{DBLP:journals/scp/AlOmarCBAONM22}, a specialized framework for mining SATD history. We replaced its original SATD detecting engine with DebtHunter to ensure methodological consistency. By applying this modified version of SATDBailiff to the commit history of our target projects, we were able to systematically identify commits where SATD was removed. We then analyzed these commits to determine if they included any modifications to the test suite. 

\smallskip\noindent\textbf{4) Linking SATD Removal to Defects.}\label{sec:bug_method}
To investigate the impact of SATD removal on software quality, we analyzed the relationship between SATD removal events and subsequent bug introductions. We first identified bug-fixing commits across the eight Java projects and employed LLM4SZZ~\cite{Tang25LLM4SZZ}, an advanced SZZ variant, to identify the specific commits and methods that introduced the bugs. 

Unlike commit-level analysis, we performed a method-level matching to ensure high precision. Specifically, we identified instances where a bug was introduced within the same method from which the SATD had been removed. This allowed us to correlate the act of debt resolution directly with the emergence of subsequent defects. Finally, we compared the linked bug-inducing rates between SATD removals accompanied by test additions and those that were not. Because SZZ can only identify bugs that were eventually detected and fixed, this rate captures defects that were observable enough to surface in the project's bug-fixing history. This comparison examines whether adding tests helps surface latent issues in these fragile areas rather than leaving them as undetected regressions.

\subsection{Data Analysis Methodology}\label{sec:data_analysis}

Our analysis follows a two-fold approach. First, we perform a cross-sectional analysis of the latest releases to measure test coverage of SATD (RQ1). We compare SATD coverage against project-level branch coverage baselines.
We then perform a longitudinal analysis of the projects' history (RQ2 and RQ3). We investigate the frequency of test additions during SATD removals (RQ2), and compare the bug introduction rates between SATD removals accompanied by tests against those that were not (RQ3), quantifying the risks associated with testless debt repayment.

\section{Results}\label{sec:results}


\newcommand{\numCommonsLangBranchPct}{92.94}
\newcommand{\numCommonsLangAuto}{6}
\newcommand{\numCommonsLangManual}{20}
\newcommand{\numCommonsLangExcluded}{2}
\newcommand{\numCommonsLangAnalyzed}{26}
\newcommand{\numCommonsLangCovered}{7}

\newcommand{\numCommonsIOBranchPct}{85.71}
\newcommand{\numCommonsIOAuto}{22}
\newcommand{\numCommonsIOMannual}{0}
\newcommand{\numCommonsIOExcluded}{1}
\newcommand{\numCommonsIOAnalyzed}{22}
\newcommand{\numCommonsIOCovered}{21}

\newcommand{\numJFreeChartBranchPct}{46.88}
\newcommand{\numJFreeChartAuto}{58}
\newcommand{\numJFreeChartManual}{2}
\newcommand{\numJFreeChartExcluded}{0}
\newcommand{\numJFreeChartAnalyzed}{60}
\newcommand{\numJFreeChartCovered}{24}

\newcommand{\numHibernateBranchPct}{58.45}
\newcommand{\numHibernateAuto}{463}
\newcommand{\numHibernateManual}{65}
\newcommand{\numHibernateExcluded}{2}
\newcommand{\numHibernateAnalyzed}{528}
\newcommand{\numHibernateCovered}{374}

\newcommand{\numSpoonBranchPct}{55.82}
\newcommand{\numSpoonAuto}{21}
\newcommand{\numSpoonManual}{4}
\newcommand{\numSpoonExcluded}{0}
\newcommand{\numSpoonAnalyzed}{25}
\newcommand{\numSpoonCovered}{13}

\newcommand{\numDubboBranchPct}{53.66}
\newcommand{\numDubboAuto}{11}
\newcommand{\numDubboManual}{0}
\newcommand{\numDubboExcluded}{10} 
\newcommand{\numDubboAnalyzed}{11}
\newcommand{\numDubboCovered}{5}

\newcommand{\numMavenBranchPct}{31.96} 
\newcommand{\numMavenLinePct}{41.73} 
\newcommand{\numMavenAuto}{55}
\newcommand{\numMavenManual}{14}
\newcommand{\numMavenExcluded}{0}
\newcommand{\numMavenAnalyzed}{69}
\newcommand{\numMavenCovered}{21}

\newcommand{\numStormBranchPct}{11.85} 
\newcommand{\numStormLinePct}{17.51} 
\newcommand{\numStormAuto}{39}
\newcommand{\numStormManual}{4}
\newcommand{\numStormExcluded}{0}
\newcommand{\numStormAnalyzed}{43}
\newcommand{\numStormCovered}{11}

\pgfmathtruncatemacro{\numTotalAuto}{%
  \numCommonsLangAuto + \numCommonsIOAuto + \numJFreeChartAuto + \numHibernateAuto + \numSpoonAuto + \numDubboAuto + \numMavenAuto + \numStormAuto
}
\pgfmathtruncatemacro{\numTotalManual}{%
  \numCommonsLangManual + \numCommonsIOMannual + \numJFreeChartManual + \numHibernateManual + \numSpoonManual + \numDubboManual + \numMavenManual + \numStormManual
}
\pgfmathtruncatemacro{\numTotalExcluded}{%
  \numCommonsLangExcluded + \numCommonsIOExcluded + \numJFreeChartExcluded + \numHibernateExcluded + \numSpoonExcluded + \numDubboExcluded + \numMavenExcluded + \numStormExcluded
}
\pgfmathtruncatemacro{\numAnalyzedSATD}{%
  \numTotalAuto + \numTotalManual
}
\pgfmathtruncatemacro{\numCoveredSATD}{%
  \numCommonsLangCovered + \numCommonsIOCovered + \numJFreeChartCovered + \numHibernateCovered + \numSpoonCovered + \numDubboCovered + \numMavenCovered + \numStormCovered
}

\newcommand{\correlationBranchAll}{0.488}
\newcommand{\correlationBranchExcludingLang}{0.901}
\newcommand{\correlationLineAll}{0.489}
\newcommand{\correlationLineExcludingLang}{0.865}


\newcommand{\numDesignAnalyzed}{505}
\newcommand{\numRequirementAnalyzed}{254}
\newcommand{\numDefectAnalyzed}{22}
\newcommand{\numTestAnalyzed}{3}
\newcommand{\numDocumentationAnalyzed}{0}

\newcommand{\numDesignCovered}{297}
\newcommand{\numRequirementCovered}{160}
\newcommand{\numDefectCovered}{16}
\newcommand{\numTestCovered}{3}
\newcommand{\numDocumentationCovered}{0}

\pgfmathtruncatemacro{\numDesignExcluded}{\numDesignSATD - \numDesignAnalyzed}
\pgfmathtruncatemacro{\numRequirementExcluded}{\numRequirementSATD - \numRequirementAnalyzed}
\pgfmathtruncatemacro{\numDefectExcluded}{\numDefectSATD - \numDefectAnalyzed}
\pgfmathtruncatemacro{\numTestExcluded}{\numTestSATD - \numTestAnalyzed}
\pgfmathtruncatemacro{\numDocumentationExcluded}{\numDocumentationSATD - \numDocumentationAnalyzed}

\newcommand{\medianDesign}{50.0}
\newcommand{\medianRequirement}{41.7} 
\newcommand{\medianDefect}{50.0}
\newcommand{\medianTest}{100}

\newcommand{\ecDA}{243}\newcommand{\ecDT}{365}
\newcommand{\ecIA}{144}\newcommand{\ecIT}{204}
\newcommand{\ecDfA}{16}\newcommand{\ecDfT}{21}
\newcommand{\ecTA}{3}\newcommand{\ecTT}{3}
\newcommand{\gpDA}{48}\newcommand{\gpDT}{116}
\newcommand{\gpIA}{15}\newcommand{\gpIT}{48}
\newcommand{\gpDfA}{0}\newcommand{\gpDfT}{1}
\newcommand{\tpDA}{6}\newcommand{\tpDT}{24}
\newcommand{\tpIA}{1}\newcommand{\tpIT}{2}


\subsection*{\rqA}\label{sec:rqa}

We applied our tracer-line instrumentation approach to all \num{\numTotalSATD} SATD instances. 
Table \ref{tab:rq1_results_per_project} presents the detailed breakdown of instrumentation process and the final test coverage results. 
First, we attempted to automatically replace the comment with an executable tracer. Instances that compiled successfully and were deemed to be in a suitable location were categorized as Automatic (B).
The remaining instances either failed the automatic step (e.g., Syntax Error) or were manually identified as being in an Inappropriate Location. We attempted to manually map these instances to a suitable executable line. 
The instances that were successfully mapped constitute the Manual (C) category. Instances that could not be mapped were Excluded (D), primarily because the SATD comment was located within a code block that was already commented out. 
Consequently, the summation of the successfully instrumented instances from the Automatic and Manual categories constitutes the final set of Analyzed SATD (E) (i.e., $E = B + C$).

\begin{table*}[hbt!]
\centering
\footnotesize
\caption{SATD Instrumentation Feasibility and Test Coverage Results per Project, Compared to Overall Project Branch Coverage.}
\label{tab:rq1_results_per_project}
\setlength{\tabcolsep}{5.5pt} 

\begin{tabular}{l ccccc S[table-format=3.1, round-mode=places, round-precision=1] S[table-format=2.1, round-mode=places, round-precision=1]} 
\toprule
 & \multicolumn{5}{c}{\textbf{Filtering Process}} & \multicolumn{2}{c}{\textbf{Coverage Result (\%)}} \\
\cmidrule(lr){2-6} \cmidrule(lr){7-8}
\textbf{Project} & 
\begin{tabular}[c]{@{}c@{}}\textbf{All}\\\textbf{SATD}\\\textbf{(A)}\end{tabular} & 
\begin{tabular}[c]{@{}c@{}}\textbf{Auto.}\\\textbf{(B)}\end{tabular} & 
\begin{tabular}[c]{@{}c@{}}\textbf{Man.}\\\textbf{(C)}\end{tabular} & 
\begin{tabular}[c]{@{}c@{}}\textbf{Excl.}\\\textbf{(D)}\end{tabular} & 
\begin{tabular}[c]{@{}c@{}}\textbf{Anal.}\\\textbf{(E)}\end{tabular} & 
{\begin{tabular}[c]{@{}c@{}}\textbf{SATD}\\\textbf{Cov.}\\\textbf{(\%)}\end{tabular}} & 
{\begin{tabular}[c]{@{}c@{}}\textbf{Overall}\\\textbf{Branch Cov.}\\\textbf{(\%)}\end{tabular}} \\ 
\midrule
\multicolumn{8}{l}{\textit{Group 1: Projects with pre-configured Jacoco plugin}} \\
\quad \lang & \num{\numCommonsLangSATD} & \num{\numCommonsLangAuto} & \num{\numCommonsLangManual} & \num{\numCommonsLangExcluded} & \num{\numCommonsLangAnalyzed} & {\calcpct{\numCommonsLangCovered}{\numCommonsLangAnalyzed}\textsuperscript{*}} & \numCommonsLangBranchPct \\
\quad \io & \num{\numCommonsIOSATD} & \num{\numCommonsIOAuto} & \num{\numCommonsIOMannual} & \num{\numCommonsIOExcluded} & \num{\numCommonsIOAnalyzed} & \calcpct{\numCommonsIOCovered}{\numCommonsIOAnalyzed} & \numCommonsIOBranchPct \\
\quad \hibernate & \num{\numHibernateSATD} & \num{\numHibernateAuto} & \num{\numHibernateManual} & \num{\numHibernateExcluded} & \num{\numHibernateAnalyzed} & \calcpct{\numHibernateCovered}{\numHibernateAnalyzed} & \numHibernateBranchPct \\
\quad \dubbo & \num{\numDubboSATD} & \num{\numDubboAuto} & \num{\numDubboManual} & \num{\numDubboExcluded} & \num{\numDubboAnalyzed} & \calcpct{\numDubboCovered}{\numDubboAnalyzed} & \numDubboBranchPct \\
\midrule
\multicolumn{8}{l}{\textit{Group 2: Projects where Jacoco plugin was manually added}} \\
\quad \jfree & \num{\numJFreeChartSATD} & \num{\numJFreeChartAuto} & \num{\numJFreeChartManual} & \num{\numJFreeChartExcluded} & \num{\numJFreeChartAnalyzed} & \calcpct{\numJFreeChartCovered}{\numJFreeChartAnalyzed} & \numJFreeChartBranchPct \\
\quad \spoon & \num{\numSpoonSATD} & \num{\numSpoonAuto} & \num{\numSpoonManual} & \num{\numSpoonExcluded} & \num{\numSpoonAnalyzed} & \calcpct{\numSpoonCovered}{\numSpoonAnalyzed} & \numSpoonBranchPct \\
\quad \maven & \num{\numMavenSATD} & \num{\numMavenAuto} & \num{\numMavenManual} & \num{\numMavenExcluded} & \num{\numMavenAnalyzed} & \calcpct{\numMavenCovered}{\numMavenAnalyzed} & \numMavenBranchPct \\
\quad \storm & \num{\numStormSATD} & \num{\numStormAuto} & \num{\numStormManual} & \num{\numStormExcluded} & \num{\numStormAnalyzed} & \calcpct{\numStormCovered}{\numStormAnalyzed} & \numStormBranchPct \\
\midrule
\textbf{Total} & 799 & 675 & 109 & 15 & 784 & {\textbf{\calcpct{\numCoveredSATD}{\numAnalyzedSATD}}} & {\textbf{---}} \\
\bottomrule
\multicolumn{8}{p{0.9\textwidth}}{\textsuperscript{*} \scriptsize This low coverage is an outlier caused by a cluster of 18 identical, untested Design SATD in Validate.java.}
\end{tabular}
\end{table*}

\medskip
The coverage results in Table \ref{tab:rq1_results_per_project} provide the first answer to our research question. 
Overall, we found that \calcpct{\numCoveredSATD}{\numAnalyzedSATD}\% (\num{\numCoveredSATD} out of \num{\numAnalyzedSATD}) of the analyzable SATD instances are covered by tests.
A deeper analysis reveals that the projects are not uniform and instead follow three distinct testing patterns. 

\smallskip\noindent\textit{1. Extra Care:} In three projects---\io, \hibernate (both Group 1), and \storm (Group 2)---we observed SATD coverage that is higher than the project's overall branch coverage baseline (\eg \io: \calcpct{\numCommonsIOCovered}{\numCommonsIOAnalyzed}\% vs. \num[round-mode=places, round-precision=1]{\numCommonsIOBranchPct}\% baseline; \storm: \calcpct{\numStormCovered}{\numStormAnalyzed}\% vs. \num[round-mode=places, round-precision=1]{\numStormBranchPct}\% baseline). This suggests that in these projects, developers pay at least equal, if not more, attention to known debt.

\smallskip\noindent\textit{2. General Postponement:} In four projects---\dubbo (from Group 1), and \jfree, \spoon, and \maven (from Group 2)---the SATD coverage was roughly proportional to, or slightly lower than, their project baselines (\eg \dubbo: \calcpct{\numDubboCovered}{\numDubboAnalyzed}\% vs. \num[round-mode=places, round-precision=1]{\numDubboBranchPct}\% baseline; \maven: \calcpct{\numMavenCovered}{\numMavenAnalyzed}\% vs. \num[round-mode=places, round-precision=1]{\numMavenBranchPct}\% baseline). The \dubbo case is particularly revealing: despite being in the Jacoco-configured group, its SATD is not prioritized, indicating that a pre-existing testing culture does not, by itself, guarantee extra care for SATD.

\smallskip\noindent\textit{3. Targeted Postponement:} \lang (from Group 1) presents a notable contrast. Despite maintaining a strict CI threshold that yields a high observed branch coverage of \num[round-mode=places, round-precision=1]{\numCommonsLangBranchPct}\%, its SATD coverage is remarkably low at\calcpct{\numCommonsLangCovered}{\numCommonsLangAnalyzed}\%.

We investigated this outlier and found it is caused by a single cluster of 18 identical Design SATD comments in the Validate.java file (see Listing \ref{fig:example-validate-before}).
The comment refers to a potential future enhancement changing a \texttt{void} method to return a \texttt{value} on the success path. 
Our instrumentation (Listing \ref{fig:example-validate-after}) confirmed that this success path is entirely untested. The project's high \num[round-mode=places, round-precision=1]{\numCommonsLangBranchPct}\% branch coverage is achieved by thoroughly testing only the failure path (the \texttt{IllegalArgumentException}).
This demonstrates that even strictly-enforced project-metric thresholds can be insufficient, as developers may focus tests on satisfying CI while leaving known technical debt on other paths untested.

\begin{lstlisting}[
    float=h!,
    language=Java,
    basicstyle=\ttfamily\small,
    commentstyle=\color{commentgreen},
    keywordstyle=\color{keywordblue},
    breaklines=true,
    breakatwhitespace=false,
    captionpos=t,
    abovecaptionskip=10pt,
    belowcaptionskip=12pt,
    caption={Example of 18 identical SATD comments in their original, inappropriate location (\lang, Validate.java).},
    label={fig:example-validate-before}
]
public static void exclusiveBetween(final double start, final double end, final double value) {
    // TODO when breaking BC, consider returning value
    if (value <= start || value >= end) {
        throw new IllegalArgumentException(String.format(DEFAULT_EXCLUSIVE_BETWEEN_EX_MESSAGE, value, start, end));
    }
}
\end{lstlisting}

\begin{lstlisting}[
    float=h!,
    language=Java,
    basicstyle=\ttfamily\small,
    commentstyle=\color{commentgreen},
    keywordstyle=\color{keywordblue},
    breaklines=true,
    breakatwhitespace=false,
    captionpos=t,
    abovecaptionskip=10pt,
    belowcaptionskip=12pt,
    caption={The code from Listing \ref{fig:example-validate-before} after manual instrumentation with the tracer line.},
    label={fig:example-validate-after}
]
public static void exclusiveBetween(final double start, final double end, final double value) {
    if (value <= start || value >= end) {
        throw new IllegalArgumentException(String.format(DEFAULT_EXCLUSIVE_BETWEEN_EX_MESSAGE, value, start, end));
    }
    System.out.println("SATD: DESIGN");
}
\end{lstlisting}



\begin{table*}[t!]
\small
\centering
\caption{Distribution of Test Coverage rates per SATD Type and Pattern.}
\label{tab:satd_intent_coverage}
\begin{tabular}{lccc}
\toprule
\textbf{Classification} & \textbf{1. Extra Care} & \textbf{2. General Post.} & \textbf{3. Targeted Post.} \\
 & (Tested / Total) & (Tested / Total) & (Tested / Total) \\
\midrule

Design & 
\multicolumn{1}{r}{\ecDA / \ecDT\ (\calcpct{\ecDA}{\ecDT}\%)} & 
\multicolumn{1}{r}{\gpDA / \gpDT\ (\calcpct{\gpDA}{\gpDT}\%)} & 
\multicolumn{1}{r}{\tpDA / \tpDT\ (\calcpct{\tpDA}{\tpDT}\%)} \\

Requirement & 
\multicolumn{1}{r}{\ecIA / \ecIT\ (\calcpct{\ecIA}{\ecIT}\%)} & 
\multicolumn{1}{r}{\gpIA / \gpIT\ (\calcpct{\gpIA}{\gpIT}\%)} & 
\multicolumn{1}{r}{\tpIA / \tpIT\ (\calcpct{\tpIA}{\tpIT}\%)} \\

Defect & 
\multicolumn{1}{r}{\ecDfA / \ecDfT\ (\calcpct{\ecDfA}{\ecDfT}\%)} & 
\multicolumn{1}{r}{\gpDfA / \gpDfT\ (\calcpct{\gpDfA}{\gpDfT}\%)} & 
\multicolumn{1}{c}{---} \\

Test & 
\multicolumn{1}{r}{\ecTA / \ecTT\ (\calcpct{\ecTA}{\ecTT}\%)} & 
\multicolumn{1}{c}{---} & 
\multicolumn{1}{c}{---} \\

\midrule
\textbf{Total} 
               & \fpeval{\ecDA+\ecIA+\ecDfA+\ecTA} / \fpeval{\ecDT+\ecIT+\ecDfT+\ecTT} \textbf{(\fpeval{round((\ecDA+\ecIA+\ecDfA+\ecTA)/(\ecDT+\ecIT+\ecDfT+\ecTT)*100,1)}\%)}
               & \fpeval{\gpDA+\gpIA+\gpDfA} / \fpeval{\gpDT+\gpIT+\gpDfT} \textbf{(\fpeval{round((\gpDA+\gpIA+\gpDfA)/(\gpDT+\gpIT+\gpDfT)*100,1)}\%)}
               & \fpeval{\tpDA+\tpIA} / \fpeval{\tpDT+\tpIT} \textbf{(\fpeval{round((\tpDA+\tpIA)/(\tpDT+\tpIT)*100,1)}\%)} \\
\bottomrule
\end{tabular}
\end{table*}


\newpage
\smallskip\noindent\textbf{Coverage by SATD Type.}
Beyond project-level patterns, we investigated whether the type of technical debt influences its likelihood of being tested.
Table \ref{tab:satd_intent_coverage} summarizes the coverage rates for each SATD type across the three project patterns. Within each pattern, the differences among SATD types are modest compared to the differences between patterns: in the Extra Care group, Design (\num{66.6}\%), Requirement (\num{70.6}\%), and Defect (\num{76.2}\%) all attain coverage above \num{66}\%, while in the General Postponement group, coverage drops to \num{41.4}\% for Design and \num{31.3}\% for Requirement with the single Defect instance left uncovered. The Targeted Postponement group records the lowest values (Design: \num{25.0}\%, Requirement: \num{50.0}\%).

Aggregating across projects yields the same picture. The median coverages for Design (\num{\medianDesign}\%), Requirement (\num{\medianRequirement}\%), and Defect (\num{\medianDefect}\%) cluster closely together with widely overlapping interquartile ranges. Contrary to the intuitive expectation that high-risk defect debt would be prioritized for testing, the technical type of SATD does not appear to influence whether it gets tested.

Test debt is a special case: all \num{\numTestAnalyzed} instances come from \hibernate and yield a median of \num{\medianTest}\%. These comments flag inadequacies in the existing tests themselves, and the surrounding code was fully exercised even though the developer considered the test logic incomplete.

\summarybox{\textbf{Answer to RQ1}}{
We found that\calcpct{\numCoveredSATD}{\numAnalyzedSATD}\% (\num{\numCoveredSATD} out of \num{\numAnalyzedSATD}) of the analyzable SATD instances are covered by tests, and there are three distinct patterns of care: (1) Extra Care in three projects, (2) General Postponement in four projects, and (3) Targeted Postponement in Commons Lang where debt-ridden paths were selectively ignored.\\
Furthermore, all three major SATD types (Design, Requirement, and Defect) show similar median coverages (clustered between \num{\medianRequirement}\% – \num{\medianDefect}\%) and widely overlapping distributions, suggesting that the technical type of SATD does not significantly influence whether it gets tested.
}

\newcommand{\cioTotal}{22}
\newcommand{\cioCo}{1}
\newcommand{\cioCoP}{4.5}
\newcommand{\cioVer}{1}
\newcommand{\cioVerP}{4.5}

\newcommand{\hibTotal}{3125}
\newcommand{\hibCo}{439}
\newcommand{\hibCoP}{14.0}
\newcommand{\hibVer}{123}
\newcommand{\hibVerP}{3.9}

\newcommand{\strTotal}{200}
\newcommand{\strCo}{7}
\newcommand{\strCoP}{3.5}
\newcommand{\strVer}{1}
\newcommand{\strVerP}{0.5}

\newcommand{\dubTotal}{323}
\newcommand{\dubCo}{127}
\newcommand{\dubCoP}{39.3}
\newcommand{\dubVer}{26}
\newcommand{\dubVerP}{8.0}

\newcommand{\jfcTotal}{173}
\newcommand{\jfcCo}{1}
\newcommand{\jfcCoP}{0.6}
\newcommand{\jfcVer}{1}
\newcommand{\jfcVerP}{0.6}

\newcommand{\mavTotal}{1188}
\newcommand{\mavCo}{37}
\newcommand{\mavCoP}{3.1}
\newcommand{\mavVer}{6}
\newcommand{\mavVerP}{0.5}

\newcommand{\spoTotal}{67}
\newcommand{\spoCo}{34}
\newcommand{\spoCoP}{50.7}
\newcommand{\spoVer}{13}
\newcommand{\spoVerP}{19.4}

\newcommand{\claTotal}{77}
\newcommand{\claCo}{3}
\newcommand{\claCoP}{3.9}
\newcommand{\claVer}{3}
\newcommand{\claVerP}{3.9}

\newcommand{\grandTotal}{5175}
\newcommand{\grandCo}{649}
\newcommand{\grandCoP}{12.5}
\newcommand{\grandVer}{174}
\newcommand{\grandVerP}{3.4}

\newcommand{\grandValid}{517}


\subsection*{\rqB}\label{sec:rqb}

Findings from RQ1 revealed that over 60\% of SATD instances are covered by existing tests in latest release revision. This suggests that code containing technical debt is generally within the reach of the project's testing infrastructure. However, from a software maintenance perspective, a critical juncture is the resolution of this debt. When an SATD comment is removed, it often signifies that the suboptimal logic has been refactored or replaced. It is therefore vital to understand whether such SATD removals trigger the creation of new tests to verify the updated implementation.

To identify commits involving SATD removals, we applied a modified version of SATDBailiff~\cite{DBLP:journals/scp/AlOmarCBAONM22} to the main branch of each project. While the original tool relies on SATD Detector, we integrated DebtHunter as the underlying engine to maintain consistency with our identification criteria in RQ1 (see Table~\ref{tab:rq2_instance_based_results} for per-project counts).

Our methodology for detecting test additions involves a logical diff of test method names between consecutive commits. First, we perform AST analysis (using \texttt{javalang\footnote{https://github.com/c2nes/javalang}}) on each file in the main branch to extract sets of method names marked with the \texttt{@Test} annotation. By comparing these sets across revisions, we identify newly added test methods. To prevent false positives caused by simple refactorings, such as method renaming, we cross-reference Git patch information to verify the similarity between added and deleted lines. Test methods identified as renames are excluded, ensuring that we only count genuine functional additions to the test suite.

In this analysis, we focus on simultaneous co-evolution within a single commit. This is supported by \textit{Wang et al.} \cite{DBLP:conf/wcre/Wang000W21}, who found that production and test code co-evolution primarily happens in the same commit (median $\approx$ 47.0\%) and that the number of co-evolutions decreases over time. Thus, same-commit analysis serves as a reliable proxy for determining whether developers add targeted tests as part of the SATD removal process.
To ensure the validity of our findings, we complemented the automated identification with a rigorous manual inspection. While we first identified \grandCo{} SATD instances where test addition and SATD removal occurred within the same commit, these automated results do not necessarily imply a causal or logical relationship between the two.

To bridge this gap, two of the authors independently inspected each of the \grandCo{} instances to determine whether the added test methods were specifically intended to exercise or verify the logic related to the removed SATD. For each instance, both inspectors examined the code changes in the commit and the content of the newly added test methods to assess their relevance to the resolved debt. The two inspectors initially agreed on the classification for \num{78.3}\% of the instances, yielding a Cohen's $\kappa$ of 0.55, which indicates moderate agreement according to the interpretation by Landis and Koch~\cite{landis1977measurement}. 
Additionally, during this initial screening, the inspectors identified and filtered out 127 false positives produced by DebtHunter (e.g., non-SATD comments incorrectly flagged as SATD) to ensure data purity. Disagreements were subsequently resolved through adjudication by a third author, who independently reviewed the conflicting cases and made the final determination. This dual-inspection protocol with third-party adjudication allows for a more precise and reproducible assessment of how often developers explicitly address technical debt through testing. The inspectors have over 5-15 years of programming experience and prior familiarity with the SATD and testing literature.

\begin{table}[t]
    \centering
    \small
    \caption{Test Addition Rates for SATD Removal Instances}
    \label{tab:rq2_instance_based_results}
    
    \newcolumntype{R}{>{\raggedleft\arraybackslash}X}
    \newcolumntype{H}{>{\centering\arraybackslash}X}
    
    \begin{tabularx}{\columnwidth}{l r R R}
        \toprule
        \textbf{Project} & \multicolumn{1}{c}{\textbf{Total SATD}} & \multicolumn{1}{H}{\textbf{w/ Co-added Tests (\%)}} & \multicolumn{1}{H}{\textbf{w/ Verified Tests (\%)}} \\
        \midrule
        \textit{Extra Care (RQ1 pattern)} \\
        ~~Hibernate    & \hibTotal   & \hibCo(\hibCoP\%)   & \hibVer(\hibVerP\%) \\
        ~~Storm        & \strTotal   & \strCo(\strCoP\%)   & \strVer(\strVerP\%) \\
        ~~Commons IO   & \cioTotal   & \cioCo(\cioCoP\%)   & \cioVer(\cioVerP\%) \\
        \midrule
        \textit{General Postponement (RQ1 pattern)} \\
        ~~Maven        & \mavTotal   & \mavCo(\mavCoP\%)   & \mavVer(\mavVerP\%) \\
        ~~Dubbo        & \dubTotal   & \dubCo(\dubCoP\%)   & \dubVer(\dubVerP\%) \\
        ~~JFreeChart   & \jfcTotal   & \jfcCo(\jfcCoP\%)   & \jfcVer(\jfcVerP\%) \\
        ~~Spoon        & \spoTotal   & \spoCo(\spoCoP\%)   & \spoVer(\spoVerP\%) \\
        \midrule
        \textit{Targeted Postponement (RQ1 pattern)} \\
        ~~Commons Lang & \claTotal   & \claCo(\claCoP\%)   & \claVer(\claVerP\%) \\
        \midrule
        \textbf{Total} & \textbf{\grandTotal} & \textbf{\grandCo(\grandCoP\%)} & \textbf{\grandVer(\grandVerP\%)} \\
        \bottomrule
        \addlinespace
        \multicolumn{4}{p{0.96\columnwidth}}{\footnotesize
            \textit{Note:}
            \textbf{Co-added Tests}: Tests introduced in the same commit as the SATD removal instance.
            \textbf{Verified Tests}: A subset of co-added tests manually confirmed to specifically target the logic of the removed SATD.
        }
    \end{tabularx}
\end{table}

\smallskip
As shown in Table~\ref{tab:rq2_instance_based_results}, our instance-level analysis reveals a trend distinct from the static snapshots in RQ1. Initially, we identified \grandCo{} instances where SATD removal and test addition co-occurred within the same commit. However, our rigorous manual inspection revealed that some of these were not genuine SATD removals, such as file renamings, misidentified comments, or instances where the SATD was unresolved.

After excluding these cases, we confirmed \grandValid{} valid co-occurring instances. Among all \num{5175} SATD removals, only \grandVer{} instances (\grandVerP\%) were verified as having new tests specifically targeting the removed SATD logic. The gap between the overall co-addition rate (\grandCoP\%) and the verified rate (\grandVerP\%) indicates that even when tests appear in the same commit as an SATD removal, most are not directed at the resolved logic. Such co-occurring tests are typically peripheral changes (touching unrelated test fixtures, fixing pre-existing test failures, or covering other features modified in the same commit) rather than targeted verification of the debt resolution.

This gap varies considerably across projects. \spoon{} and \dubbo{} show the largest drops, with \spoon{} falling from \spoCoP\% co-added to \spoVerP\% verified and \dubbo{} from \dubCoP\% to \dubVerP\%. In these two projects, developers frequently bundle test changes with SATD removal commits, but most of those tests target unrelated logic.  \hibernate{}, despite contributing the largest absolute number of verified instances (\hibVer{}), shows a low per-removal verified rate of \hibVerP\%, suggesting that even projects with extensive testing infrastructure rarely add targeted verification at the moment of debt resolution.


\summarybox{\textbf{Answer to RQ2}}{
While SATD instances are generally covered by existing tests (RQ1), our manual verification reveals that only \grandVerP{}\% of SATD removals are accompanied by new tests specifically tailored to the debt resolution. This suggests that SATD removal is predominantly performed without targeted verification, relying instead on a pre-existing safety net that may not be specifically optimized for the updated implementation.
}


\newcommand{\claBF}{424}\newcommand{\claBI}{263}
\newcommand{\cioBF}{232}\newcommand{\cioBI}{108}
\newcommand{\hibBF}{3199}\newcommand{\hibBI}{2300}
\newcommand{\dubBF}{1503}\newcommand{\dubBI}{1074}
\newcommand{\spoBF}{1152}\newcommand{\spoBI}{977}
\newcommand{\mavBF}{1030}\newcommand{\mavBI}{820}
\newcommand{\strBF}{717}\newcommand{\strBI}{590}
\newcommand{\jfcBF}{74}\newcommand{\jfcBI}{46}

\newcommand{\sumBF}{\fpeval{\claBF + \cioBF + \hibBF + \dubBF + \spoBF + \mavBF + \strBF + \jfcBF}}
\newcommand{\sumBI}{\fpeval{\claBI + \cioBI + \hibBI + \dubBI + \spoBI + \mavBI + \strBI + \jfcBI}}

\newcommand{\numTotalBICommits}{10933}
\newcommand{\numTotalBuggyStatements}{132572}
\newcommand{\pctSOneSuccess}{58.5}
\newcommand{\pctSTwoSuccess}{59.4}
\pgfmathsetmacro{\pctBothSuccess}{46.4}

\newcommand{\rqcTotalTest}{174}
\newcommand{\rqcTotalNo}{4874}

\newcommand{\bicThirtyTest}{3}\newcommand{\bicThirtyTestP}{1.72}
\newcommand{\bicThirtyNo}{92}\newcommand{\bicThirtyNoP}{1.89}

\newcommand{\bicNinetyTest}{4}\newcommand{\bicNinetyTestP}{2.30}
\newcommand{\bicNinetyNo}{115}\newcommand{\bicNinetyNoP}{2.36}

\newcommand{\bicOneEightyTest}{5}\newcommand{\bicOneEightyTestP}{2.87}
\newcommand{\bicOneEightyNo}{141}\newcommand{\bicOneEightyNoP}{2.89}

\newcommand{\bicThreeSixtyTest}{7}\newcommand{\bicThreeSixtyTestP}{4.02}
\newcommand{\bicThreeSixtyNo}{163}\newcommand{\bicThreeSixtyNoP}{3.34}

\newcommand{\bicNoneTest}{11}\newcommand{\bicNoneTestP}{6.32}
\newcommand{\bicNoneNo}{213}\newcommand{\bicNoneNoP}{4.37}

\newcommand{\bicDayZeroTest}{2}\newcommand{\bicDayZeroTestP}{1.15}
\newcommand{\bicDayZeroNo}{61}\newcommand{\bicDayZeroNoP}{1.25}

\newcommand{\medianDaysTest}{456}
\newcommand{\medianDaysNo}{198}

\newcommand{\churnTest}{297.0}
\newcommand{\churnNo}{236.0}
\newcommand{\linesAddTest}{234.0}
\newcommand{\linesAddNo}{130.0}
\newcommand{\methodSizeTest}{26.0}
\newcommand{\methodSizeNo}{24.0}

\subsection*{\rqC}\label{sec:rqc}

To establish a foundation for identifying defects potentially introduced by SATD removal, we first characterized the bug dataset collected using LLM4SZZ. As shown in Table~\ref{tab:bug_dataset_stats}, we analyzed a total of \num{\sumBF} bug-fixing commits across eight open-source Java projects. From these, our pipeline successfully identified at least one bug-inducing commit for \num{\sumBI} bug-fixes, yielding an overall identification rate of\calcpct{\sumBI}{\sumBF}\%. This high success rate provides a sufficient corpus for tracing defects back to their origins and examining their proximity to SATD removal events.

\begin{table}[t]
    \centering
    \caption{Statistics of Bug-Inducing Commit Identification via LLM4SZZ}
    \label{tab:bug_dataset_stats}
    \begin{tabular}{lrrr}
        \toprule
        Project & \begin{tabular}[t]{@{}r@{}}Bug-Fixing \\ Commits\end{tabular} & \begin{tabular}[t]{@{}r@{}}With Bug- \\ Inducing\end{tabular} & \begin{tabular}[t]{@{}r@{}}Rate\end{tabular} \\ \midrule
        \lang  & \num{\claBF}   & \num{\claBI}   & \calcpct{\claBI}{\claBF}\% \\
        \io    & \num{\cioBF}   & \num{\cioBI}   & \calcpct{\cioBI}{\cioBF}\% \\
        \hibernate & \num{\hibBF} & \num{\hibBI} & \calcpct{\hibBI}{\hibBF}\% \\
        \dubbo         & \num{\dubBF} & \num{\dubBI} & \calcpct{\dubBI}{\dubBF}\% \\
        \spoon         & \num{\spoBF} & \num{\spoBI} & \calcpct{\spoBI}{\spoBF}\% \\
        \maven         & \num{\mavBF} & \num{\mavBI} & \calcpct{\mavBI}{\mavBF}\% \\
        \storm         & \num{\strBF}   & \num{\strBI}   & \calcpct{\strBI}{\strBF}\% \\
        \jfree    & \num{\jfcBF}    & \num{\jfcBI}    & \calcpct{\jfcBI}{\jfcBF}\% \\ \midrule
        \textbf{Total} & \textbf{\num{\sumBF}} & \textbf{\num{\sumBI}} & \textbf{\calcpct{\sumBI}{\sumBF}}\% \\ \bottomrule
    \end{tabular}
\end{table}

The identification process utilized two complementary strategies: direct Large Language Model (LLM) analysis of patch content and a combination of traditional SZZ with LLM-based validation. The direct analysis (S1) succeeded in \num{\pctSOneSuccess}\% of the files, while the SZZ+LLM approach (S2) succeeded in \num{\pctSTwoSuccess}\%. Both strategies converged on the same result in \num{\pctBothSuccess}\% of the cases, demonstrating the consistency of our methodology. In total, we identified \num{\numTotalBICommits} unique bug-inducing commits, linked to \num{\numTotalBuggyStatements} individual buggy statements.

Our analysis in RQ2 established that only 3.4\% of all SATD removals (\num{\grandVer} instances) are manually validated as being targeted by dedicated test modifications. To assess whether this scarcity influences subsequent defects, we compared bug induction rates between the test-added and no-test groups across multiple observation windows, excluding the 127 false positives filtered out during the RQ2 manual validation. Table~\ref{tab:bug_lifecycle_analysis} summarizes the results. 
Across short-to-medium-term intervals (30, 90, and 180 days), the bug induction rates are nearly identical between the two groups (\bicThirtyTestP\% vs. \bicThirtyNoP\%, \bicNinetyTestP\% vs. \bicNinetyNoP\%, and \bicOneEightyTestP\% vs. \bicOneEightyNoP\%, respectively). A slight divergence becomes visible over longer windows, with the test-added group reaching \bicNoneTestP\% versus \bicNoneNoP\% when no time limit is imposed. 

These results are counterintuitive: one might expect tests added during debt resolution to suppress subsequent defects. The bugs counted here, however, are those eventually identified and fixed in the same method as the SATD removal, not necessarily defects caused by the resolution itself, which complicates causal interpretation. Several factors may contribute to the observed pattern. The accompanying tests may not have been sufficient to verify the resolved logic. Developers may also add tests selectively to SATDs that are particularly critical or complex, whose inherent difficulty produces defects despite the presence of tests. Disentangling these mechanisms is beyond the scope of this study and warrants future investigation, possibly motivating the development of SATD-aware test prioritization techniques.

\begin{table}[htbp]
    \centering
    \small
\caption{Bug-Inducing Commits Linked to SATD Removals across Observation Windows}    \label{tab:bug_lifecycle_analysis}
    \newcolumntype{C}{>{\centering\arraybackslash}X}
    \begin{tabularx}{\columnwidth}{l C C}
        \toprule
        \textbf{Metric} & \textbf{Tests Added} & \textbf{No Tests} \\
        \midrule
        Within 30 days  & \bicThirtyTest{} (\bicThirtyTestP\%)  & \bicThirtyNo{} (\bicThirtyNoP\%) \\
        Within 90 days  & \bicNinetyTest{} (\bicNinetyTestP\%)  & \bicNinetyNo{} (\bicNinetyNoP\%) \\
        Within 180 days & \bicOneEightyTest{} (\bicOneEightyTestP\%)  & \bicOneEightyNo{} (\bicOneEightyNoP\%) \\
        No time limit   & \bicNoneTest{} (\bicNoneTestP\%)  & \bicNoneNo{} (\bicNoneNoP\%) \\
        \bottomrule
    \end{tabularx}
\end{table}


\summarybox{\textbf{Answer to RQ3}}{
SATD resolutions accompanied by test modifications show no immediate reduction in linked bug induction rates within short-to-medium-term windows (\bicThirtyTestP\% vs. \bicThirtyNoP\% within 30 days), and a slight long-term gap (\bicNoneTestP\% vs. \bicNoneNoP\% with no time limit) is difficult to attribute to a single cause. The observed pattern may reflect insufficient test targeting, selection effects toward particularly complex SATDs, or tests surfacing latent defects that would otherwise go unrecorded. 
}

\section{Discussion and Implications}\label{sec:implications}
This section reflects on the three results presented above and considers
what they imply for SATD management and software testing practice.
We first interpret the patterns observed in test coverage, co-evolution,
and defect alignment, and then derive implications for researchers
and practitioners.

\smallskip
\noindent
\textbf{Interpretation of Results.} The results suggest that the relationship between technical debt and testing is not straightforward. While one might expect developers to be more careful with code they know is suboptimal, our analysis in RQ1 indicates that this only happens in a minority of cases. Most projects follow a pattern of General Postponement, where technical debt is tested no more (and sometimes less) than the rest of the codebase. This suggests that the same factors leading to the introduction of technical debt, such as time pressure, likely also lead to insufficient testing of that code.

A particularly interesting observation is the Targeted Postponement pattern, exemplified by the Commons Lang project. In this case, the project maintains very high branch coverage, yet the specific execution paths containing technical debt are completely untested. This happens because tests are focused on ``failure paths'' (e.g., verifying that exceptions are thrown) to satisfy coverage metrics, while the ``success paths'' where the technical debt resides are ignored. This finding suggests that high-level coverage metrics can provide a false sense of security, as they do not necessarily reflect the testing quality of the most risk-prone parts of the system.

The longitudinal analysis in RQ2 and RQ3 reveals a disconnect in how testing accompanies the SATD lifecycle. Developers rarely synchronize test modifications with debt resolution, and the long-term differences in linked bug rates between the test-added and no-test groups admit multiple non-exclusive interpretations: tests may be insufficient to verify the resolved logic, developers may add tests selectively to particularly complex SATDs, or tests may surface latent defects that would otherwise go unrecorded. While we cannot adjudicate among these interpretations within the present data, they converge on a common implication: SATD repayment is currently treated as an ordinary code change rather than a high-risk maintenance activity warranting dedicated verification.

\noindent
\textbf{Implications.}
Our results converge on the implication that SATD repayment requires more deliberate verification than it currently receives. This finding has several implications for both researchers and practitioners.

For researchers, our study suggests several directions for future work. Since tests accompanying SATD removal may be insufficient to exercise the resolved logic, researchers should develop SATD-aware test generation techniques that produce tests targeting the specific code paths involved in the debt. Because developers also appear to add tests selectively, often to particularly complex SATDs, SATD prioritization techniques that identify which debt items most warrant verification effort would help focus testing resources. Furthermore, the impact of CI/CD coverage thresholds on test quality warrants further investigation: our findings suggest that high global coverage does not necessarily translate to a robust safety net, and strict thresholds may incentivize developers to create tests merely to satisfy numerical requirements rather than to verify complex debt-ridden areas.

For practitioners, these findings offer practical guidance for debt management. Project managers should not rely solely on project-wide coverage percentages, which can mask the lack of validation in critical, debt-heavy areas. Instead, they should supplement these with local coverage analysis of code marked with \texttt{TODO} or \texttt{FIXME} tags. Furthermore, organizations should treat SATD resolution as a high-risk maintenance activity warranting dedicated verification, rather than relying on existing safety nets that may not be adequate for the resolved logic.

\section{Related Work}\label{sec:relatedwork}

\subsection{Empirical Studies of Testing}
Large-scale empirical studies have consistently found a weak positive correlation between test volume and defect counts across open-source projects.
\textit{Kochhar et al.}~\cite{DBLP:conf/qsic/KochharBLJ13} examined over 20,000 open-source projects, finding that projects with more code and more developers tend to have more test cases, and identifying a weak positive correlation between test volume and defect counts.
\textit{Islam et al.}~\cite{DBLP:conf/msr/IslamHBH23} revisited these findings in a decade-long longitudinal study of Java projects, confirming that while testing grows alongside the codebase, this weak correlation remained stable through 2021.
Beyond static metrics, \textit{Beller et al.}~\cite{DBLP:journals/tse/BellerGPPAZ19} used IDE instrumentation to show that testing is frequently treated as an afterthought rather than a proactive activity.

The timing of test changes relative to production code has also been studied: while \textit{Marsavina et al.}~\cite{DBLP:conf/scam/MarsavinaRZ14} established fundamental co-evolution patterns, \textit{Wang et al.}~\cite{DBLP:conf/wcre/Wang000W21} found that co-evolutions predominantly occur in the same commit (median $\approx$ 47\%) and that their frequency decreases over time.
Comparative studies on TDD vs.\ test-last development~\cite{DBLP:journals/tse/FucciETOJ17, DBLP:conf/ispw/TosunATJ18} further show that different methodologies yield distinct coverage profiles: TDD tends to encourage higher branch coverage through refined test cases, while test-last development results in broader method-level coverage.
Despite these insights, existing literature rarely addresses whether these practices apply specifically to code explicitly marked as technical debt, which is the gap our work fills.

\subsection{Self-Admitted Technical Debt}
Since \textit{Potdar and Shihab}~\cite{DBLP:conf/icsm/PotdarS14} formally introduced SATD, showing it appears in 2.4\%--31\% of source files with 26--64\% eventually removed, substantial work has accumulated on identification and classification.
Task tags (\texttt{TODO}, \texttt{FIXME}, \texttt{XXX}) have been established as reliable SATD signals~\cite{DBLP:journals/tosem/GuoLLLCLZ21}; complementary efforts include ontologies standardizing debt vocabulary~\cite{DBLP:conf/icsm/AlvesRCMS14}, ML-based methods for distinguishing design from requirement debt~\cite{DBLP:journals/tse/MaldonadoST17}, and LLM-based approaches for fine-grained classification~\cite{DBLP:journals/ese/SheikhaeiTWX24}.
Regarding lifecycle and impact, \textit{Wehaibi et al.}~\cite{DBLP:conf/wcre/WehaibiSG16} showed that while SATD-related changes do not always induce more immediate defects, they are more complex and difficult to perform, thereby hindering future modifications; \textit{Mensah et al.}~\cite{DBLP:journals/jss/MensahKSBM18} further identified inadequate testing (21.2\%) as one of the leading drivers of SATD, directly motivating our study.

\textit{Maldonado et al.}~\cite{DBLP:conf/icsm/MaldonadoASS17} found that the majority of SATD is eventually resolved, frequently by the same developer who introduced it, persisting between 82--613 days on average before removal.
To manage the resulting volume, \textit{Lima et al.}~\cite{DBLP:journals/sqj/LimaGE22} proposed a hybrid prioritization approach combining SATD descriptions with static analysis findings, achieving greater precision than comment-text-only methods.
Despite these advances, while \textit{Mensah et al.} identified inadequate testing as a \emph{cause} of SATD, there remains a lack of empirical evidence on how testing practices are employed to safeguard the repayment process, which is the focus of this research.

\section{Threats to Validity}\label{sec:threats_to_validity}


\noindent\textbf{Construct validity} concerns whether our metrics and procedures accurately capture the phenomena we intended to study. One potential threat to construct validity involves our use of tracer-line instrumentation as a proxy for measuring the test coverage of SATD instances. While inserting an executable statement directly into the debt-ridden code block provides a precise way to verify execution, it does not guarantee that the tests fully exercised the specific logic described in the technical debt comment. For example, a tracer line confirms that a code block was reached, but not necessarily that all execution sub-paths within that block were covered. We mitigated this by manually mapping tracers to the most relevant lines within the debt-ridden block when automatic insertion was not possible.

Another threat related to construct validity is our method for linking SATD removal to subsequent defects. Our primary analysis imposes no time limit, treating any bug introduced in the same method as potentially related to the SATD removal. We report results at 30, 90, and 180-day windows as sensitivity analyses. While the no-limit approach is most conservative about excluding relevant bugs, it may also include bugs introduced by unrelated changes to the same method. Additionally, we cannot fully rule out that the higher bug-detection rate in the test-added group is partially explained by greater change complexity; a controlled analysis matching commits by churn size would be needed to isolate the effect of test addition from the effect of commit scale.


\noindent\textbf{Internal validity} refers to the extent to which the observed results are due to the factors under study rather than confounding variables. The accuracy of the automated tools used in our study, such as DebtHunter, SATDBailiff, and LLM4SZZ, represents a potential threat to internal validity. To mitigate errors in SATD detection, we manually validated the results on a subset of our projects and excluded identified false positives. For the identification of bug-inducing commits, we used LLM4SZZ, which employs large language models that can exhibit stochastic behavior. We addressed this by using its dual-strategy approach and confirming the consistency between independent identification strategies.

There is also a risk of human error during the manual mapping of tracer lines for complex cases, such as inappropriate comment locations or syntax errors. To minimize this risk, we followed a protocol where instrumentation was carefully reviewed to ensure that the tracer line accurately reflected the execution of the logic referred to by the SATD comment, rather than just the literal position of the original comment.


\noindent\textbf{External validity} concerns the generalizability of our findings to other contexts. We analyzed eight open-source Java projects from different domains and of varying scales. However, our findings may not generalize to commercial software, projects written in other programming languages, or systems with different testing cultures. We tried to broaden the scope of our observations by including projects with both pre-existing and manually integrated testing infrastructures.

The scope of the SATD instances we analyzed also presents a threat to generalizability. Our study was strictly limited to technical debt found within method bodies, which was necessary to measure executable coverage precisely. This means we excluded technical debt documented at the class or package level, such as architectural debt. Consequently, our conclusions primarily apply to technical debt related to implementation and local design rather than high-level system architecture.


To ensure the reliability and replicability of our study, we relied on established open-source tools including Jacoco, DebtHunter, and SATDBailiff. We have also provided a comprehensive replication package that contains our datasets, instrumentation scripts, and analysis results to allow other researchers to verify and build upon our work.

\section{Conclusion}\label{sec:conclusion}
We empirically investigated the relationship between SATDs and software testing in eight open-source Java projects, examining 784 SATD instances for cross-sectional test coverage and 5,175 SATD removal events for longitudinal co-evolution and defect alignment.

Our central finding is a co-evolution gap between debt resolution and test maintenance. Although 60.7\% of SATD-affected code is covered by existing tests, only 3.4\% of SATD removals are accompanied by tests specifically targeting the resolved debt (vs.\ 12.5\% that co-add any tests). Developers, therefore, treat SATD repayment as an ordinary code change rather than as a high-risk maintenance activity, despite their own prior flag.

Moreover, these rare test additions show no evidence of preventing short- to medium-term regressions: bug induction rates within 30, 90, and 180 days are nearly identical across the test-added and no-test groups (\eg 1.72\% vs.\ 1.89\% within 30 days). This null effect need not mean tests are ineffective in general; developers may selectively add tests to particularly difficult SATDs, whose complexity dominates the outcome. Separating test effectiveness from the difficulty of the underlying debt, and designing SATD-aware verification that prioritizes the hardest cases, are promising future directions.


\section{Acknowledgment}
We gratefully acknowledge the financial support of JSPS KAKENHI grants (JP24K02921, JP25K03100, JP26K23818, JP26H02500), as well as ASPIRE grant (JPMJAP2415), JST CREST grant (JPMJCR23M1, JPMJCR26X7), and JST FOREST (JPMJFR252W).

\section{Data Availability}\label{sec:DataAvailability}
To support the reproducibility of our study, we have made our collection pipelines publicly available as replication packages. 
Throughout this research, we comprehensively utilized two repositories to build our foundational pipeline: the customized version of DebtHunter used for method-level SATD detection and precise metadata extraction~\cite{artifact_debthunter}, and the history mining framework that integrates our customized DebtHunter engine into SATDBailiff to trace SATD-removing events throughout the project lifecycle~\cite{artifact_satdbailiff}. 
For the long-term defect analysis in RQ3, the replication package containing our execution scripts and environment for defect tracking and LLM4SZZ~\cite{Tang25LLM4SZZ} analysis is provided at~\cite{artifact_bug_analysis}.



\balance
\bibliography{references.bib} 


\end{document}